\documentclass[%
groupedaddress,
eunsortedaddress,
 preprint,
 showpacs,
 amsmath,amssymb,
 aps,
 showkeys,
]{revtex4-1}

\usepackage{graphicx}
\usepackage{dcolumn}
\usepackage{bm}
\usepackage[makeroom]{cancel}

\newcommand{\ang}{\text \AA}
\newcommand{\angs}{\text \AA $~$}
\newcommand{\kunit}{$\tfrac{\text{kcal}}{\text{mol} \cdot \ang^2}~$}
\newcommand{\sod}{$\text{Na}^+~$}
\newcommand{\pot}{$\text K^+~$}

\begin{document}

\preprint{APS/123-QED}

\title{Stiff-spring approximation revisited: inertial effects in non-equilibrium trajectories}

\author{Mostafa Nategholeslam}
\email[seyed.mostafa@gmail.com]{}
\affiliation{Department of Physics, University of Guelph, Guelph, Ontario, Canada}

\author{C. G. Gray}
\email[cgray@uoguelph.ca]{}
\affiliation{Guelph-Waterloo Physics Institute and Department of Physics, University of Guelph, Guelph, Ontario, Canada}
\affiliation{Department of Physics and Biophysics Interdepartmental Group, University of Guelph, Guelph, Ontario, Canada}

\author{Bruno Tomberli}
\email[brunotomberli@capilanou.ca]{}
\affiliation{Department of Physics, Capilano University, North Vancouver, British Columbia, Canada}
\affiliation{Department of Physics and Biophysics Interdepartmental Group, University of Guelph, Guelph, Ontario, Canada}

\date{\today}

\begin{abstract}
Use of harmonic guiding potentials is perhaps the most commonly adopted method for implementing steered molecular dynamics (SMD) simulations, performed to obtain potentials of mean force (PMFs) of molecular systems using Jarzynski's equality and other non-equilibrium work (NEW) theorems.
Harmonic guiding potentials are also the natural choice in single molecule force spectroscopy experiments such as optical tweezers and atomic force microscopy, performed to find the potential of mean force using NEW theorems.
The stiff spring approximation (SSA) of Schulten and coworkers enables to use the work performed along many SMD trajectories in Jarzynski's equality to obtain the PMF.

We discuss and demonstrate how a high spring constant, $k$, required for the validity of the SSA can violate another requirement of this theory, i.e., the validity of Brownian dynamics of the system under study, if the value of $k$ is too high.
Violation of the Brownian condition results in the introduction of kinetic energy contributions to the external work, performed during SMD simulations.
These \textit{inertial effects} result in skewed work distributions, rather than the Gaussian distributions predicted by SSA.
The inertial effects also result in broader work distributions, which in turn worsen the effect of the skewness when one tries to calculate reliable work averages.
Remarkably, neither the skewness nor the broadening of work distributions can be attributed to factors other than using too-stiff springs.
In particular, our results strongly suggest that the skew and width of work distributions are independent of the average drift velocity and the asymmetries of the physical systems studied, at least for the range of systems and velocities we examined.

The skew and broadening of work distributions result in biased estimation of the underlying PMF, using Jarzynski's equality or the more efficient forward-reverse (FR) method of Kosztin and coworkers.
This pathology is more pronounced in larger biomolecular simulations, where longer samplings are required to achieve convergence.
The bias manifests in such simulations in the form of a systematic error that increases with simulation time.
We discuss the proper upper limit for $k$, such that the inertial effects are practically avoided.
Used together with the relation for the lower limit of $k$ (which follows naturally from considering thermal fluctuations per degree of freedom of the system), the practitioner of SMD can then conduct accurate steering, while satisfying
the Brownian dynamics requirements of SSA.
Even in cases where inertial effects (due to high $k$) result in biased estimations of the PMF, we argue and demonstrate that using the peak-value (rather than the statistical mean) of the work distributions vastly reduces the bias in the calculated PMFs and improves the accuracy.

\end{abstract}

\maketitle

\section{Introduction}
Among the many attempts to use the Jarzynski equality \cite{Jarzynski1,Jarzynski2} for obtaining free energy profiles in molecular-level experiments and simulations, the stiff spring approximation (SSA)\cite{SMD1,SMD2,SMD3} has been tremendously successful.
This has been mostly due to the role SSA plays as the theoretical basis for the steered molecular dynamics (SMD) method.
Due to its conceptual simplicity and ease of implementation, SMD has been widely used in studying biomolecular processes\cite{FR,OFR,SMD1,SMD2,Jarzynski_Application_1,Jarzynski_Application_2,Jarzynski_Application_3,Jarzynski_Application_4,Jarzynski_Application_5,Jarzynski_Application_6,Jarzynski_Application_7,Jarzynski_Application_8,Jarzynski_Application_9,Jarzynski_Application_10,Jarzynski_Application_11,Jarzynski_Application_12,Jarzynski_Application_13,Jarzynski_Application_14,Jarzynski_Application_15,Jarzynski_Application_16,Jarzynski_Application_17,Jarzynski_Application_18,Jarzynski_Application_19}.
SSA provides a comprehensive theoretical platform for SMD simulations, by elucidating  the relation between the harmonic guiding potentials and the behavior and distribution of non-equilibrium work performed during such simulations \cite{SMD1,SMD2,SMD3}.
One remarkable use of SSA came through its invocation by Kosztin and coworkers in devising the forward-reverse (FR) method \cite{FR}, which has provided an even simpler mechanism for both understanding the non-equilibrium work distributions in SMD simulations and using them to obtain the PMFs for systems under study.

Jarzynski's equality \cite{Jarzynski1,Jarzynski2} gives the free energy difference $\Delta \Phi_{AB}$ between two macrostates $A$ and $B$ of a many-body system, as being equal to logarithm of the exponential average of the external work on the system, $W_{A\rightarrow B}$, performed over many distinct micropaths that evolve the system from $A$ to $B$,
\begin{equation}
  e^{-\beta \Delta \Phi_{AB}} = \langle e^{-\beta W_{A\rightarrow B}} \rangle
\label{Jarzynski}
\end{equation}
where $\beta \equiv (k_B T)^{-1}$ and $\langle...\rangle$ denotes (path) ensemble averaging.
Macrostates $A$ and $B$ are characterized by the values $x_A$ and $x_B$ of a reaction coordinate $x$, which is often an externally controllable parameter of the system, such as the relative distance between two chosen molecules.
The thermodynamic state parameters of the system should be well determined at state $A$, but they can deviate from their initial values at $A$, during the non-equilibrium evolution of the system to state $B$.
SSA asserts that when the external work in Jarzynski's equality is performed using a \textit{sufficiently stiff} spring, its distribution among an ensemble of trajectories will be Gaussian.
Marcinkiewicz's theorem \cite{Marcinkiewicz} then implies that the cumulant expansion of the logarithm of the right-hand-side of Jarzynski's equality (\ref{Jarzynski}) can be safely terminated at the second order.
This fact has been utilized by many authors in extracting free energy profiles, or PMFs, from non-equilibrium SMD simulations (see e.g. \cite{Jarzynski_Application_7,Jarzynski_Application_13}), where the first and second cumulants of work are used to obtain the free energy difference between the state $A$ and the subsequent macrostates along the chosen range of the reaction path.

This already provides a better means of estimating $\Delta \Phi$, compared to a direct application of Jarzynski's equality; finding the first two cumulants of work distribution i.e., $\langle W \rangle$ and $\langle W^2 \rangle - \langle W \rangle^2$ from a finite ensemble of trajectories has a much higher rate of convergence (to a pre-specified accuracy) than sampling the exponential of external work $\langle \exp (-\beta W) \rangle$ \cite{SMD3}.
The convergence of the exponential average depends heavily on the small and even negative values of $W$, whose occurrence is rare by the second law of thermodynamics.

The FR method takes this one step further, by performing the steering process in both forward and reverse directions.
This simple change dramatically improves the convergence rate to the underlying PMF, by liberating the simulator even from sampling the second cumulant of work distribution.
The underlying mechanism can be best seen when external work in forward ($W_F$) and reverse ($W_R$) directions are written as the sum of their reversible and irreversible parts.
The reversible part of the forward(reverse) work is by definition equal to $\Delta \Phi_{AB}$ ($\Delta \Phi_{BA} = -\Delta \Phi_{AB}$).
When steering in both directions is done using the same constant average speed, the irreversible or dissipative portion of the forward and reverse works are (most often) equal on average, i.e. $\langle W^{diss} \rangle_{F} = \langle W^{diss} \rangle_{R}$.
Here $F$ (forward) refers to $A \rightarrow B$ process and $R$ (reverse) refers to the $A \leftarrow B$ process.
Possible exceptions might include Brownian steering of an asymmetrical object along a translational reaction path, while the orientation of the object is held fixed.
However, in most cases, the average forward and reverse dissipative works are equal.
In particular, it has been shown that when SSA requirements are met, this equality holds \cite{FRApp1}.

As the main task of NEW theorems is to decompose the external work into reversible ($W^{rev}$) and dissipative ($W^{diss}$) parts, one can exploit the equality of $\langle W^{diss} \rangle$ in forward and reverse directions and write

\begin{align}
\Delta \Phi_{AB} &= \langle W^{rev} \rangle_{F} = -\langle W^{rev} \rangle_{R}
= \frac{\langle W^{rev} \rangle_{F} -\langle W^{rev} \rangle_{R}}{2} = \nonumber \\
&= \frac{(\langle W \rangle_F - \langle W^{diss} \rangle_{F}) - (\langle W \rangle_R - \langle W^{diss} \rangle_{R})}{2}\nonumber \\
& = \frac{\langle W \rangle_F - \langle W \rangle_R}{2}
\label{FR}
\end{align}
This gives $\Delta \Phi_{AB}$ simply as half the difference of forward and reverse works, and the average dissipative work can then be calculated as $\langle W \rangle_F -\Delta \Phi_{AB}~(=\langle W \rangle_R +\Delta \Phi_{AB}) $.

SSA formally puts no upper limit on the stiffness of the spring used for steering the systems under study, and only requires the harmonic potential to be sufficiently strong (stiff).
In particular, to conduct accurate steering one requires that $k \gg max\{|d^2\Phi/dx^2|\}$, for $x_A \leq x \leq x_B$.
This is to guarantee the desired accuracy in steering the system along the prescribed path and also to guarantee that the distribution of the work values among separate trajectories is Gaussian \cite{SMD3}.
While this is an essential requirement, the profile of $\Phi(x)$ is in general a priori unknown (and hence it's first and second derivatives), so one often uses a criteria such as that given in Eq.\ref{k low}, for the lower bound on the chosen value for $k$.

SSA is most often invoked in SMD simulations, wherein time is discretized into \textit{time-steps} of finite size $\delta t$.
The motion of the steered object then follows a discrete path, jumping one time-step movement at a time.
During each time-step, a steering force $-k[x_{current}-x_{target}]$ is applied to each steered object, aiming it at its respective desired location by the end of the time-step, as prescribed by the simulator.
If $k$ is not sufficiently large, the steered object will not closely follow the prescribed trajectory $x_{target}(t)$, as the external force applied during most time-steps will not be sufficiently large to steer the object against the underlying potential of the system and thermal fluctuations.
On the other hand, if $k$ is too large, one will most often apply too large a steering force on the object and \textit{overshoot} it from the $x_{target}$ prescribed for the object by the end of the current time-step.
This artifact occurs only in discretized trajectories and can result in deviation of the distribution of external work from a Gaussian shape, which subsequently leads to biased estimation of the PMF through NEW theorems.
This bias has previously been reported by other investigators \cite{Jarzynski_bias1,Jarzynski_bias2}, and attempts have been made to formulate and isolate the error \cite{Jarzynski_bias2}.
Here we argue and demonstrate that this error originates in the choice of the spring constant.

Let us assume that for a given system, there exists an optimal value $k_O$ for the stiffness of the steering potential, which results in minimum average deviation of the steered object from the prescribed path, over the course of an SMD simulation.
Using any $k < k_O$ results in loose, inaccurate steering, and $k > k_O$ results in the overshooting phenomenon just described.
In the latter case, every time that the steered object overshoots its target position, an extra amount of external work is performed on the system, which has not been spent on overcoming the local PMF barrier or thermal fluctuations, but rather on giving extra kinetic energy to the steered object.
Although this extra kinetic energy will be absorbed by the surrounding heat reservoir (or any computational mechanism resembling its effect, e.g., the temperature control algorithms in molecular dynamics simulations), its trace will remain in the recorded external work performed by the restraint.
One should recall that SSA builds upon the assumption of the system being in the overdamped (i.e. Brownian) limit of the Langevin equation.
In a system with the PMF $\Phi(x)$, subject to a harmonic guiding potential $\tfrac{k}{2} (x-vt)^2$, the overdamped Langevin equation reads
\begin{equation}
\frac{dx}{dt} = - \beta D(x) \frac{\partial }{\partial x} \left(\Phi(x) + \frac{k}{2} (x-vt)^2 \right) + \sqrt{2 D(x)}~ \xi ,
\label{Langevin}
\end{equation}
where $\xi$ is the white-noise variable satisfying $\langle \xi(t) \xi(t') \rangle = \delta(t - t')$ and $D(x)$ is the local diffusion coefficient.
There is no acceleration term in (\ref{Langevin}), as in the Brownian limit the $m d^2x/dt^2$ term is overwhelmed by drag and random forces, and thus dropped from the Langevin equation.
This will not hold if $k$ is chosen to be so large that the $\tfrac{k}{2} (x-vt)^2$ term can accelerate the steered objects.
In that limit, one has to replace the Brownian particle model (\ref{Langevin}) with the more general Brownian oscillator model, where the $m d^2x/dt^2$ is also present.
The former model serves as a basis for establishing the SSA, while the latter model cannot easily serve the same purpose.
In the absence of discretization, this problem would be far less serious.
Choice of the value of $k$ should thus be made with two criteria in mind: $k$ should be large enough to enable precise steering, but shall be small enough to let (\ref{Langevin}) remain valid.
If the latter requirement is not satisfied in an SMD simulation, in attempting to force the particle to achieve a constant target velocity, v, by discretized corrections with an average momentum correction of zero, large $-k  \Delta x$ forces will accelerate the steered objects.
This gives larger magnitude contributions to recorded values of external work above the mean (since $[(v + \delta v)^2 - v^2] \ge [v^2 -(v -\delta v)^2]$), which result in work distributions that are skewed toward higher values of energy.
This occurs both in forward and reverse work samplings, although in general not to the same extent.
When input to either the Jarzynski equality or the FR method (\ref{FR}), this right-skewness of work distributions will then result in systematic error, or bias, in the calculated PMFs.
Such a bias has been reported by other authors \cite{Jarzynski_bias1,Jarzynski_bias2}, and its relation to the underlying work distributions is investigated in the following sections.
In a recent publication \cite{bin-passing}  we have presented an efficient method for implementing the FR method, in the absence of these inertial effects.
In the following, we first demonstrate the outcome of the inertial effects just described on the characteristics of non-equilibrium work distributions and then on the PMF estimations obtained using (the average of) those work distributions.
We then provide the criteria one needs to satisfy in designing SMD simulations so that the inertial effects are avoided or minimized.
A comprehensive scheme is then provided for extracting the PMFs using (\ref{FR}), wherein the peak value of the work distributions is used instead of the work average.
We provide physical justification for this choice and then demonstrate how this method (which we call the {\it peak-finding} method) even partially avoids the bias when inertial effects are present and improves the accuracy of the PMF calculation in large (bio)molecular simulations.
Our peak-finding method involves two stages.
First the work distributions are built from the data of the applied force and the resultant displacements along SMD trajectories.
In the next stage, the peak-value of these distributions are found using a linear curve-fitting procedure, and used to calculate the PMF.
We provide software for two separate algorithms, each performing one stage of this process, both under terms of the GNU general public license, version 3.
The peak-finding method presented here is based on the same principal idea that forms the basis of the bin-passing method \cite{bin-passing}: when analyzing an SMD trajectory, external work values from individual time-steps can be considered separately and used towards calculating the average work function in the proper direction, based on actual direction of progress of the system during each time-step.
This is in contrast to the more widely used {\it bin-crossing} scheme, where work readings from successive time-steps are averaged together, regardless of the fluctuations in actual direction of progress of the system from one time-step to the next.

\section{Alteration of external work distributions by inertial effects}

Although one formally requires a stiff-enough spring to reliably perform the steering in SMD simulations, such steering is inherently imperfect due to thermal fluctuations in the system, and the (a priori unknown) PMF governing the average motion along the reaction path.
When the spring constant $k$ chosen to perform the steering is too large, the overshooting phenomenon described in the introduction will happen often, and results in work values (in both directions) much larger than the typical residual PMF differences along the reaction path.
This results in broadening the work distributions, an also in skewing them toward larger values.

\begin{figure*}[!]
\centering
\includegraphics[width=0.50\linewidth]{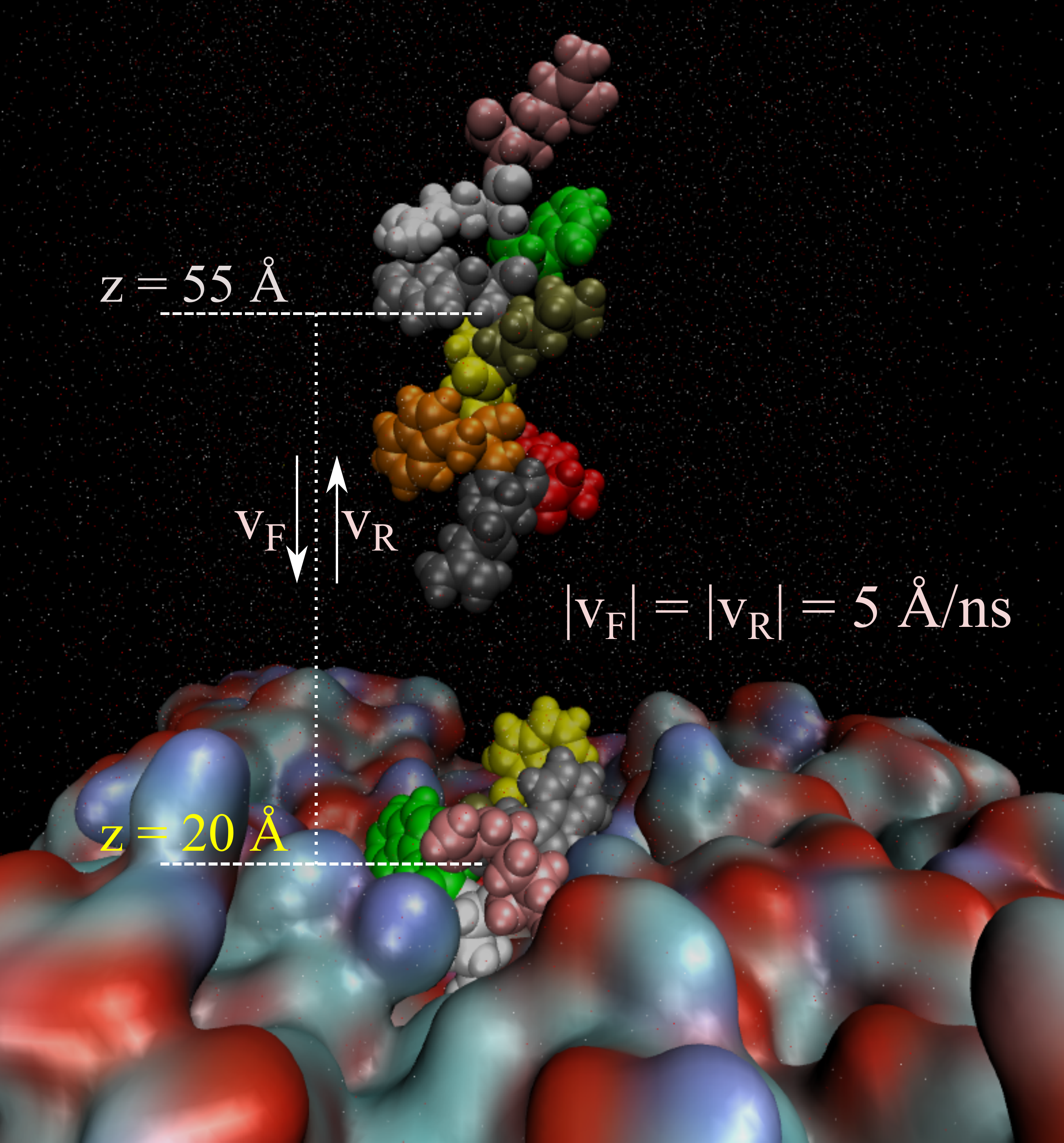}
\caption{Reaction path for an HHC-36 peptide molecule (sequence KRWWKWWRR) as it is steered to go towards the surface of a patch of mixed (3:1) POPE/POPG membrane.
The patch is composed of 128 phospholipid molecules, with 16 POPG and 48 POPE molecules in each leaflet.
Water containing 0.15 M concentration of NaCl adjoins both faces of the membrane (13531 water molecules), with the number of ions adjusted (18 \sod and 23 $\text{Cl}^-$ ions) such that the system as a whole becomes electrostatically neutral.
The dimensions of the box are approximately 78.7 \angs $\times$ 73.3 \angs $\times$ 155.6 \ang.
Simulations are all performed with NAMD \cite{NAMD} version 2.8, under NPT (P = 1.00 atm and T = 310 K) conditions, using TIP3P water model \cite{TIP3P}.
A time-step of 2 fs is used for this and all other simulations reported in this paper.
The membrane surface is shown in a coarse-grained representation here for a clearer view, but in the simulations the actual membrane patch is represented in all-atom details.
The symbols $\text{v}_F$ and $\text{v}_R$ denote the constant steering velocities in the forward and reverse directions, respectively.
Water molecules and ions are also shown as very small dots here to allow viewing the peptide and the membrane surface.
The CHARMM36 force-field \cite{CHARMM36} is used for the membrane and CHARMM27 \cite{CHARMM27} for everything else in the simulation box.}
\label{peptide-membrane_picture}
\end{figure*}

\begin{figure*}[!]
\centering
\includegraphics[width=0.90\linewidth]{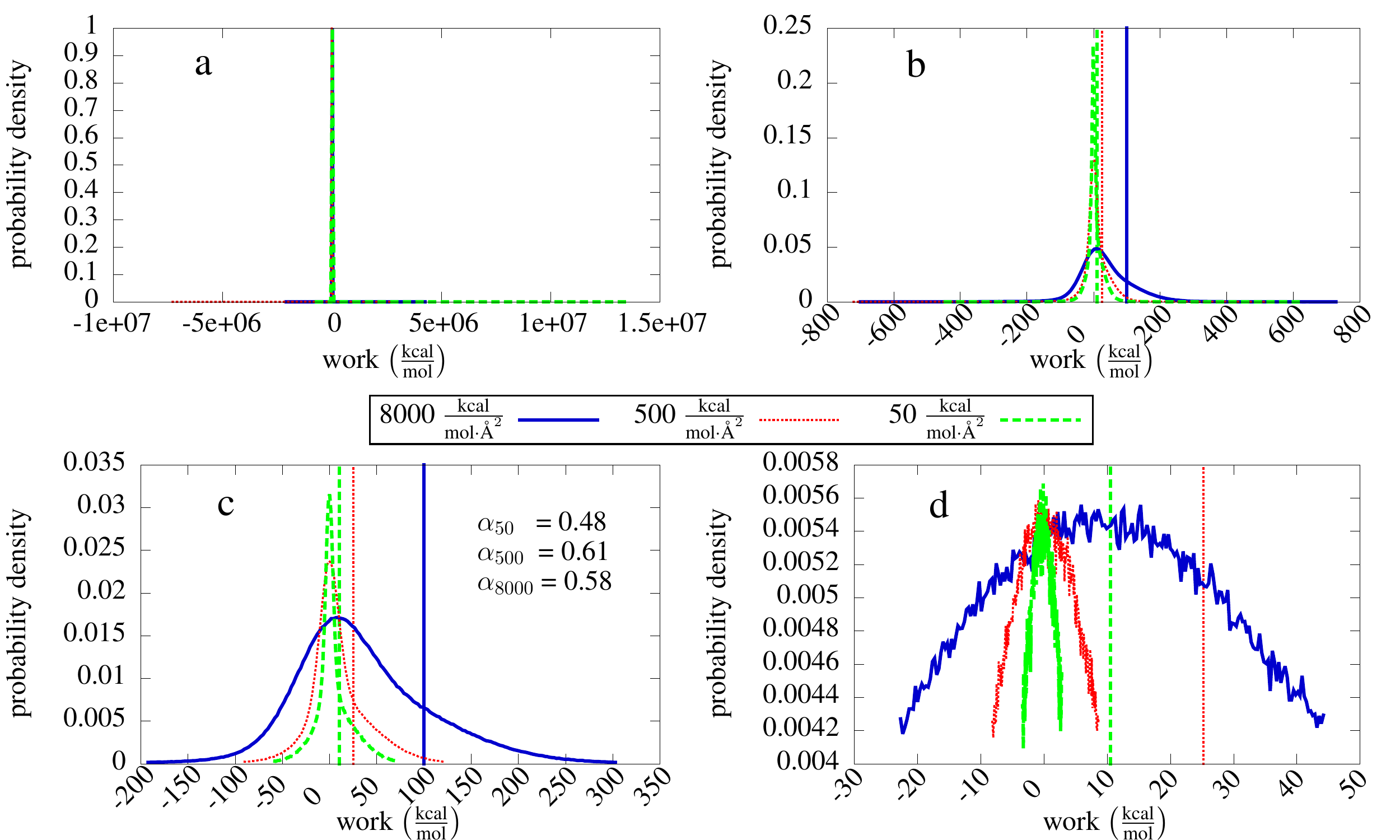}
\caption{Work distributions obtained from SMD simulations with an HHC-36 peptide steered to go toward a POPE/POPG membrane patch, using three different steering spring constants, but similar simulation conditions and steering parameters otherwise.
The distributions shown are for the forward (toward membrane) works in the bin with $z\in[30\ang,30.5\ang]$.
Work distributions from other bins are qualitatively similar to those shown here.
Further simulation details are given in the text.
Each shown work distribution is obtained by dividing the given range of work values into 200 equal bins, establishing the sample counts within each bin and then dividing the counts for each bin by the total counts in all bins, so that the area under each distribution curve equals unity.
In (a), no zooming is performed and work distributions are depicted with their respective full ranges.
This results in no visible difference between the three work distributions in (a).
In (b), (c) and (d), we have zoomed into the central region of the distribution in (a), using zooming factors ($f$) of 0.001, 0.01 and 0.75, respectively (see text).
The difference in width of the distributions, as well as the right-skewness of the distributions obtained from simulations using stiffer springs is most vivid in (b) and (c).
The difference in locations of the peaks is best seen in (d).
Average work values obtained for each spring constant (from the whole data samples, shown in (a)) are depicted in (b), (c) and (d) by vertical lines using the same color and line-style as the distribution curves with the respective spring constants.
In all cases, it is seen that $\langle W \rangle$ is larger than $W_{peak}$, and the deviation increases consistently with the stiffness of the harmonic potential employed for steering, to the extent that $\langle W \rangle$ for $k$ = 8000 \kunit  falls outside the range used in (d).}
\label{fig_peptide-membrane_distributions}
\end{figure*}

An example of such skewed work distributions is given in Fig. \ref{fig_peptide-membrane_distributions} and the resultant biased PMFs (obtained using the FR method, with the bin-passing technique \cite{PhD_thesis,bin-passing}) are shown in Figs. \ref{fig_PMFs_bin-passing_far_range} and \ref{fig_PMFs_bin-passing} for a series of FR SMD simulations performed on a peptide-membrane system.
In each of these simulations, a small peptide is steered to go towards the surface of a phospholipid membrane patch (from $z = 55$ \angs to $z = 20$ \ang) and back, with a speed of 5 \ang/ns, where $z$ is the direction normal to the surface of the membrane.
The z-component of the center of mass of the membrane is restrained to stay at zero.
More details about the composition of the peptide-membrane system and the simulations are given in the caption to Fig. \ref{peptide-membrane_picture}.
For simulations using spring constants of 8000, 500 and 50 \kunit, total simulation times are 800, 743.87 and 507.27 ns, respectively.
Each FR cycle in these series of simulations takes 15 ns, and the given total simulation times thus translate to about 53, 49 and 34 statistically independent FR cycles, respectively.
\subsection*{The notion of scaled work}
To obtain the work distributions depicted in Fig.  \ref{fig_peptide-membrane_distributions}, the whole dataset of work values for all the time-steps spent within the given bin ($z\in[30\ang,30.5\ang]$) is read for each curve.
Each single time-step work reading is scaled up according to the linear approximation inherent to all PMF calculation methods: it is assumed that the value of the PMF difference, and thus the value of the external work required to overcome the PMF barrier, is constant across each bin along the reaction path.
If the bin-width is $\Delta x$ and the system has progressed by a distance $\delta x$ along the reaction path during a given time-step, while an external work $\delta W$ is performed on the system, the scaled work will be equal to $\delta W \times \Delta x / \delta x$.
Using this notion of scaled work enables us to maximally exploit the statistics of time-steps, toward separating forward and reverse work distributions for each bin.
We hereafter refer to these scaled work values simply as work values.

The range of work values for each dataset is then divided into 200 work-bins (not to be confused with reaction coordinate bins), and populations of various work values within each work-bin are then established.
In Fig. \ref{fig_peptide-membrane_distributions}.a, the whole range of work values obtained using each spring constant is shown.
\subsection*{The zooming factor}
Typical to distributions with very wide ranges of variation of the variable under study, Fig. \ref{fig_peptide-membrane_distributions}.a provides no utility for distinguishing the features of the three shown curves from each other.
To overcome this problem, we {\it zoom} into the center of each distribution,considering work values that differ in magnitude by no more than a fraction $f$ from the peak-value of the original distribution.
We refer to $f$ as the {\it zooming factor} from this point on.
The same number of work-bins (here, 200) is used for establishing these zoomed-in work distributions, resulting in much narrower ranges of variation for work, when values of 0.001, 0.01 and 0.75 are used for $f$ in parts (b), (c) and (d) of Fig. \ref{fig_peptide-membrane_distributions}, respectively.
Average work values $\langle W \rangle$ obtained from the entire dataset of the simulation with each spring constant for the given bin are shown with vertical lines (with the same color and line style as the corresponding work distribution curve) in parts (b), (c) and (d) of Fig. \ref{fig_peptide-membrane_distributions}.
\subsection*{Skewness of the work distributions}
Parts (b) and (c) of Fig. \ref{fig_peptide-membrane_distributions} vividly exhibit that the work distribution obtained from $k = $ 8000 \kunit simulation is a skewed Gaussian, rather than the symmetric one expected from the SSA.
The skewness values ($\alpha$) reported in Figs. \ref{fig_peptide-membrane_distributions} and \ref{fig_K-Na_distributions} are calculated using the relation
\begin{equation}
  \alpha = \frac{\mu_3}{\sigma^3},
\label{skewness}
\end{equation}
wherein $\mu_3$ is the third moment around the mean (of work), calculated as
\begin{equation}
  \mu_3 = \langle (W - \langle W \rangle)^3 \rangle,
\label{mu3}
\end{equation}
where $\langle W \rangle \equiv (1/N) \sum W_i$ is the average for N work samplings $W_i$, and $\sigma$ is the standard deviation of work.

The smallest spring constant results in smallest (nearly negligible) skewness.
The width (or dispersion) of work distributions also reduces consistently with $k$.
The combined effect of reduced skewness and dispersion in work distributions is more accurate and less biased estimate of $\langle W \rangle$.
This, in turn leads to better estimation of the PMF, using (\ref{FR}), as seen in Figs. \ref{fig_PMFs_bin-passing_far_range} and \ref{fig_PMFs_bin-passing}.
\subsection*{Using $W_{peak}$ instead of $\langle W \rangle$ in the FR formulae}
SSA predicts that the distribution of work values obtained from SMD trajectories in a given system will be a (non-skewed) Gaussian, provided that $k$ is high enough.
The FR method \cite{FR,FRApp1} further implies that SMD trajectories obtained using the same steering speed should yield identical average work values for the same system.
Together, these two assertions imply that the peak location of the work distribution ($W_{peak}$) obtained using a finite steering speed under SSA simulation conditions should be the same as $\langle W \rangle$ from the same set of simulations.
One can thus use $W_{peak}$ instead of $\langle W \rangle$ in (\ref{FR}) to calculate the PMF.
This substitution may seem unnecessary and Computationally more expensive.
But use of the scaled work notion, described above, makes it economic to establish the work distributions from fairly short simulations. 
What is more, we show (e.g. in Fig. \ref{fig_PMFs_bin-passing_far_range}, that using $W_{peak}$ in place of $\langle W \rangle$ in (\ref{FR}) results in faster convergence (and with reduced bias) to the underlying PMF.
Given the whole profile of a skewed Gaussian distribution, we can find the peak location more accurately than we can calculate the average, and estimated peak location is less affected by increased skewness.
The benefits of substituting $W_{peak}$ for $\langle W \rangle$ in (\ref{FR}) are a direct result of this simple fact.

Equality of the average and the peak value of work distributions among sets of SMD simulations in a given system should be independent of the stiffness of the spring used to perform the steering, as long as $k$ is large enough to satisfy the SSA requirements \cite{SMD2,SMD3}, as the FR method does not require that different samplings be done using the same $k$.
Figs. \ref{fig_peptide-membrane_distributions} and \ref{fig_K-Na_distributions} (and many similar distributions for other reaction path bins, not shown here) manifestly shows that this does not hold;
using different spring constants to perform SMD on the same system and with the same speed has resulted in work distributions with different average and peak values of non-equilibrium work.
One nevertheless observes (Fig. \ref{fig_peptide-membrane_distributions}.d) that the peak-values of these work distributions are much closer together than their corresponding average values.

\subsection*{Effect of physical asymmetries and drift speed on the work distributions}

One possible source of the skewness observed in the work distributions is the physical asymmetries in the system under study.
As a thought experiment, we consider the classic example of a gas enclosed in a cylinder, its volume being controlled externally via the position of a piston.
We now consider a series of forward (compressing) and reverse (decompressing) events, where the volume of the gas is changed from an initial value $V_A$ to a final value $V_B < V_A$.
The compressing and decompressing events are performed at an arbitrary but constant and equal speed.
Other thermodynamical details of the process are not of user concern here, as long as we assume that the initial states of all the compressing and decompressing events are equilibrium states with the same temperature $T$.
Assuming that the initial equilibrium pressure of the system at $A$ ($P_A$, associated with $V_A$) is higher than or equal to the pressure of the surrounding environment (or heat reservoir), the forward (compressing) events would require higher work values, than the reverse (decompressing) events.
This, however, does not necessarily imply that the distribution of forward works should be skewed.
The consequence of the asymmetry in the process might simply be that the average forward work is higher than the average reverse work for this experiment, and this indeed should be the case for (\ref{FR}) to give the correct free energy difference between states $A$ and $B$.
As yet another possible source of the skewness in work distributions, we imagine various sets of these compressing-decompressing events, where successively higher drift speeds are used.
The drift speed is equal within each set.
We can now ask whether realizations of the process performed at higher drift speeds might produce skewed work distributions, as dissipative work generally increases with the drift speed, making larger (positive) contributions to the external work.

\begin{figure*}[!]
\centering
\includegraphics[width=0.60\linewidth]{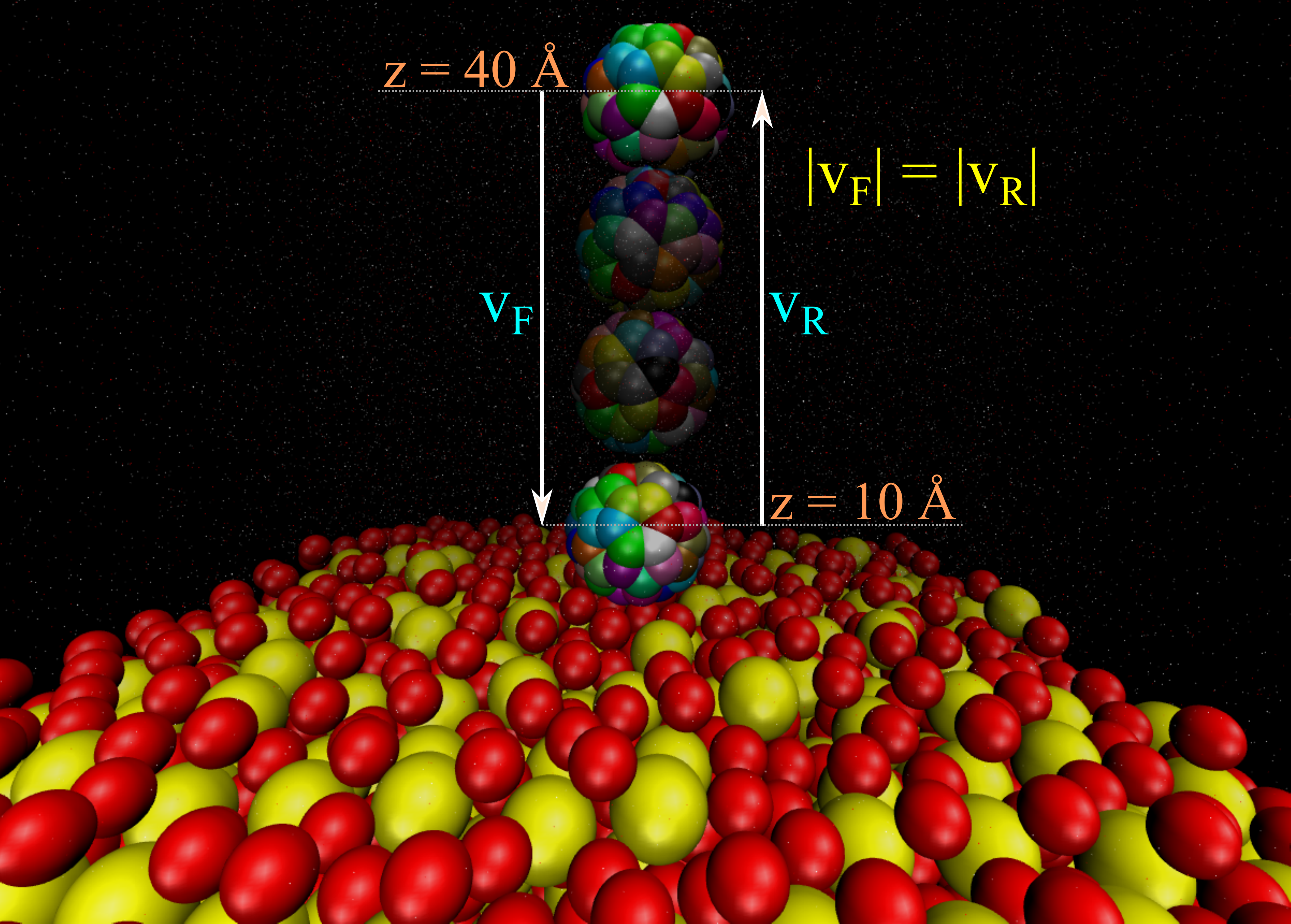}
\caption{Depiction of the reaction path for a $\text{C}_{60}$ buckminsterfullerene molecule, as it is restrained to go toward a silicon dioxide ($\text{SiO}_2$) slab and back.
The silicon dioxide (silica) slab is composed of 432 silicon and 1152 oxygen ions.
The mismatch between the ratio of silicon to oxygen ions in the system and their correct molar ratio in silica is due to finite size of the slab in crystal form used here.
The silicon and oxygen ions are ordered in successive layers, and picking a small piece of the crystal often results in getting slightly disproportionate numbers of the two types of ions.
The silica slab serves only as a rigid wall (ions are strictly fixed in place, i.e., constrained) for the $\text{C}_{60}$ molecule to interact with, and the exact molar ratio of silicon to oxygen ions in the slab is thus irrelevant.
13346 TIP3P water molecules are used to hydrate the system, resulting in a box with approximate dimensions of 72.6 \angs $\times$ 72.5 \angs $\times$ 86.5 \ang.
The silica slab extends in the xy plane, and the $\text{C}_{60}$ molecule moves back and forth along the z-axis.
The thickness of the silica slab extends roughly between z = $\pm$7.1 \ang.
With a radius of over 4 \ang, the $\text{C}_{60}$ molecule thus makes contact with the upper surface of the silica slab, when its center of mass is at its minimum z-value of 10 \ang.
Different steering speeds are used to study the speed-dependence of the shape of work distributions, as explained in the text.
Simulations are performed using NAMD (veresion 2.10) \cite{NAMD} using the CHARMM27 force field \cite{CHARMM27}, at constant pressure (P = 1.00 atm) and temperature (T = 300 K), with a time-step of 2 fs.}
\label{fig_C60-silica_picture}
\end{figure*}

To investigate these two possibilities, we consider a very asymmetric molecular system, with a $\text{C}_{60}$ buckminsterfullerene molecule being restrained to approach a silicon dioxide (silica) slab, make contact with the silica slab, and then move away from the slab.
The reaction path is the normal distance between the center of mass of the $\text{C}_{60}$ molecule and the surface of the silica slab.
The process is performed in an aqueous environment at atmospheric pressure and room temperature (T = 300 K).
More details about the simulation box are given in the caption of Fig. \ref{fig_C60-silica_picture}.
When the $\text{C}_{60}$ molecule approaches the silica slab at a finite speed, water molecules are momentarily trapped between the two objects, resembling the gas compression process explained above.
When moving away from the silica slab, the $\text{C}_{60}$ molecule experiences no similar effect with water molecules, i.e., it interacts with water molecules whose movement is not restricted by any physical barriers.

\begin{figure*}[!]
\centering
\includegraphics[width=0.90\linewidth]{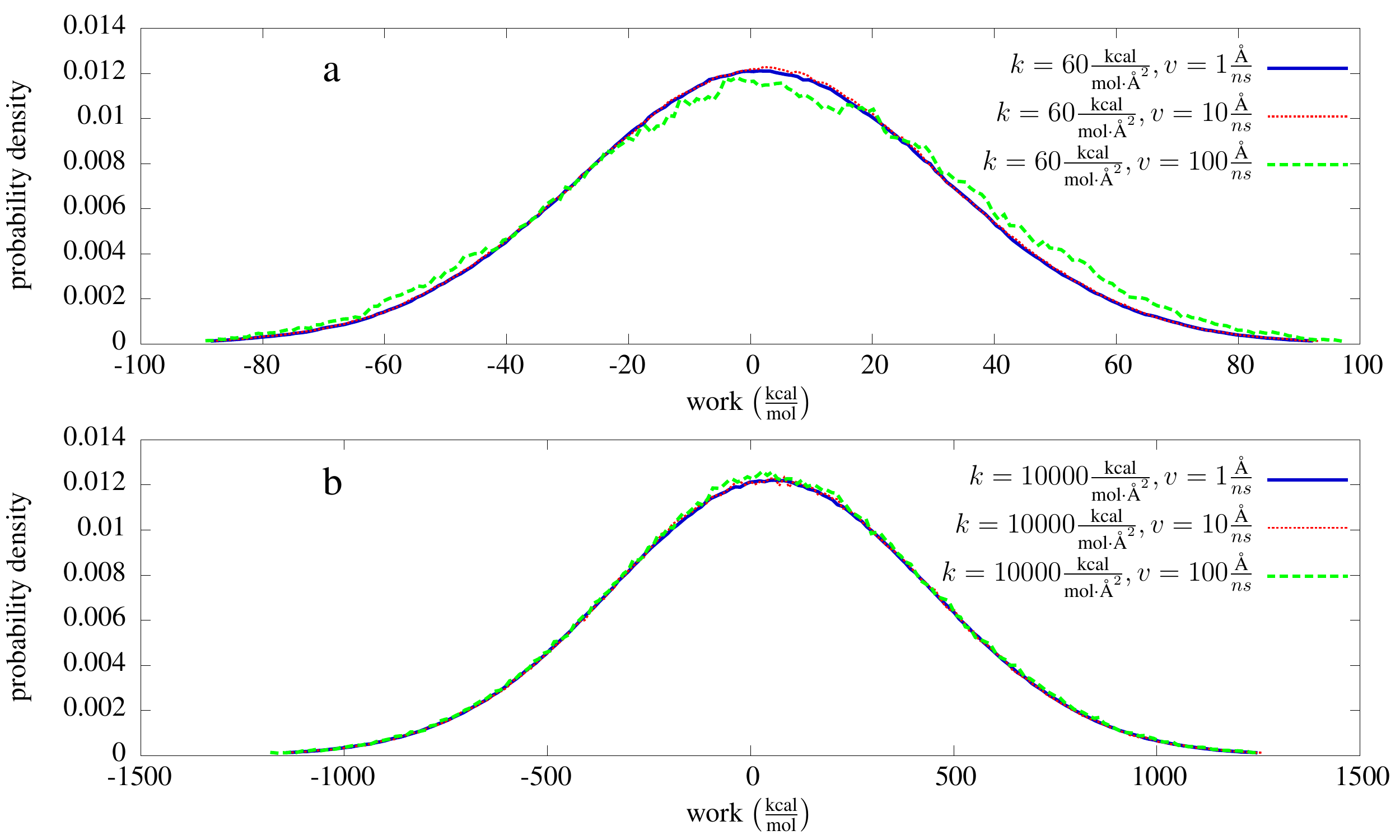}
\caption{Forward work distributions ($\text{C}_{60}$ going toward the silica slab) in the bin with $z\in[20\ang,25\ang]$, for the system shown in Fig. \ref{fig_C60-silica_picture}.
In (a), work distributions are shown from three simulations where a spring of stiffness 60 \kunit is used to steer the $\text{C}_{60}$ molecule, but with three different steering (drift) speeds of 1, 10 and 100 \ang/ns.
The distributions shown in (b) are from simulations with $k$ = 10,000 \kunit, with the same three average steering speeds used in (a).
All of the distributions in (a) and (b) are established with a zooming factor ($f$) of 0.01.
It is vividly seen in both (a) and (b) that variations in drift speed among simulations with the same steering spring constant have no significant effect on the width or skewness of the distributions.
Change of spring constant from 60 \kunit in (a) to 10,000 \kunit in (b), however, is seen to have dramatically increased the width of the work distribution, among simulations done with the same drift speed.}
\label{fig_C60-silica_distributions}
\end{figure*}
Work distributions shown in Fig. \ref{fig_C60-silica_distributions} for two different spring constants, each with three different drift speeds (six simulation series in total) show essentially no trace of skewness.
Absence of skewness even in work distributions obtained with a very high spring constant of 10,000 \kunit in Fig. \ref{fig_C60-silica_distributions}, can be attributed to constraining (rather than restraining) the silica slab.
Following the terminology of the NAMD literature \cite{NAMD}, a {\it constrained} object is one that is held strictly fixed in space, with absolutely no movement allowed for any of its atoms.
As we will explain below, external work distributions become skewed only when both objects that are steered to change position relative to one another, are {\it restrained} using very stiff springs (i.e., unlike a constrained particle, some small movement is allowed).
The absence of the restraining forces (and hence of external work) on one of the steered objects (the silica slab) here allows us to study the isolated effect of other possible causes of skewness, i.e., physical asymmetry and drift speeds.

Work distributions from simulations with drift speeds varying within two orders of magnitude and spring constants differing by three orders of magnitude, shown in Fig. \ref{fig_C60-silica_distributions}.a and b, show essentially no skewness.
These distributions are from a rather large bin ($z\in[20\ang,25\ang]$) where the distance between the $\text{C}_{60}$ molecule and the silica slab is small enough that the trapping of water molecules between the two objects is plausible.
These non-skewed work distributions (and similar ones from other bins, not shown here), strongly suggest that the skewness of work distributions can be attributed to neither physical asymmetry in the system, nor the (high) steering speeds.
Higher width of work distributions in part b of Fig. \ref{fig_C60-silica_distributions} compared to part a, is another clear example of broadening of the work distributions as a result of using stiffer steering springs.
It is important to note that the drift speed is quite different from the step-wise speed, with the latter resembling the instantaneous speed for the discretized time-steps of finite width in the simulations.
SMD simulations with higher steering $k$ generally have higher step-wise speeds and broader step-wise speed distributions, which correlates well with their broader work distributions and higher $\langle W \rangle$.
We demonstrated (e.g. in Fig. \ref{fig_C60-silica_distributions}) that the width of work distributions (from SMD runs with a given $k$) is independent of the drift (i.e., average) speeds.
Fig. \ref{fig_C60-silica_distributions} clearly shows that the width of work distributions depends strongly on $k$, and we assert that this is due to the effect of $k$ on the width and average magnitude of the distribution of step-wise speeds.
In the conventional bin-crossing method, obtaining similar work distributions is much costlier, and the width of such distributions will in general depend on the average pulling speed $v$.
\subsection*{Dependence of skewness on the mass and size of the steered objects}
The skewness of work distributions obtained from an SMD simulation with a given spring constant and drift velocity may, in general, depend also on the geometry and mass of the steered objects.
This can be studied by examining the work distributions from a series of FR SMD simulations where the distance between a sodium and a potassium ion is changed with a constant drift velocity.
The reaction coordinate here is the distance between the centers of masses of the two ions.
The potassium ion (\pot) is restrained at the origin of the coordinate system using a harmonic potential, while the sodium ion (\sod) is steered to move with the constant average velocity of 8.00 \ang/ns toward the \pot ion from a distance of 10.1 \angs to a distance of 2.1 \angs along the x-axis.
The \sod ion is then held fixed there for a duration of 0.1 ns to let the system equilibrate, and then moved back to its initial location at 10.1 \angs from the \pot ion with the same average speed of 8.00 \ang/ns.
After another 0.1 ns of equilibration, this FR cycle is started over.
The whole cycle takes 2.2 ns, and movements of the \sod ion are all along the x-axis.
In each simulation, the same spring constant is used to steer both objects, i.e., to keep the \pot fixed in place and to move the \sod along the x-axis or to hold it stationary during the equilibration intervals.
The system is hydrated with 3916 TIP3P \cite{TIP3P} water molecules, making a nearly cubical box of dimensions 50 \angs on each side.
Simulations are performed using NAMD (versions 2.8 or 2.10) \cite{NAMD}) with the CHARMM27 force field \cite{CHARMM27}, at constant temperature of 310 K, and using a time-step of 2 fs.
Initial configurations for the simulations are equilibrated states where pressure was set at 1.00 atm.
The production FR runs where performed under constant volume conditions, with the volume of the box fixed at its initial equilibrated value.

\begin{figure*}[!]
\centering
\includegraphics[width=1.00\linewidth]{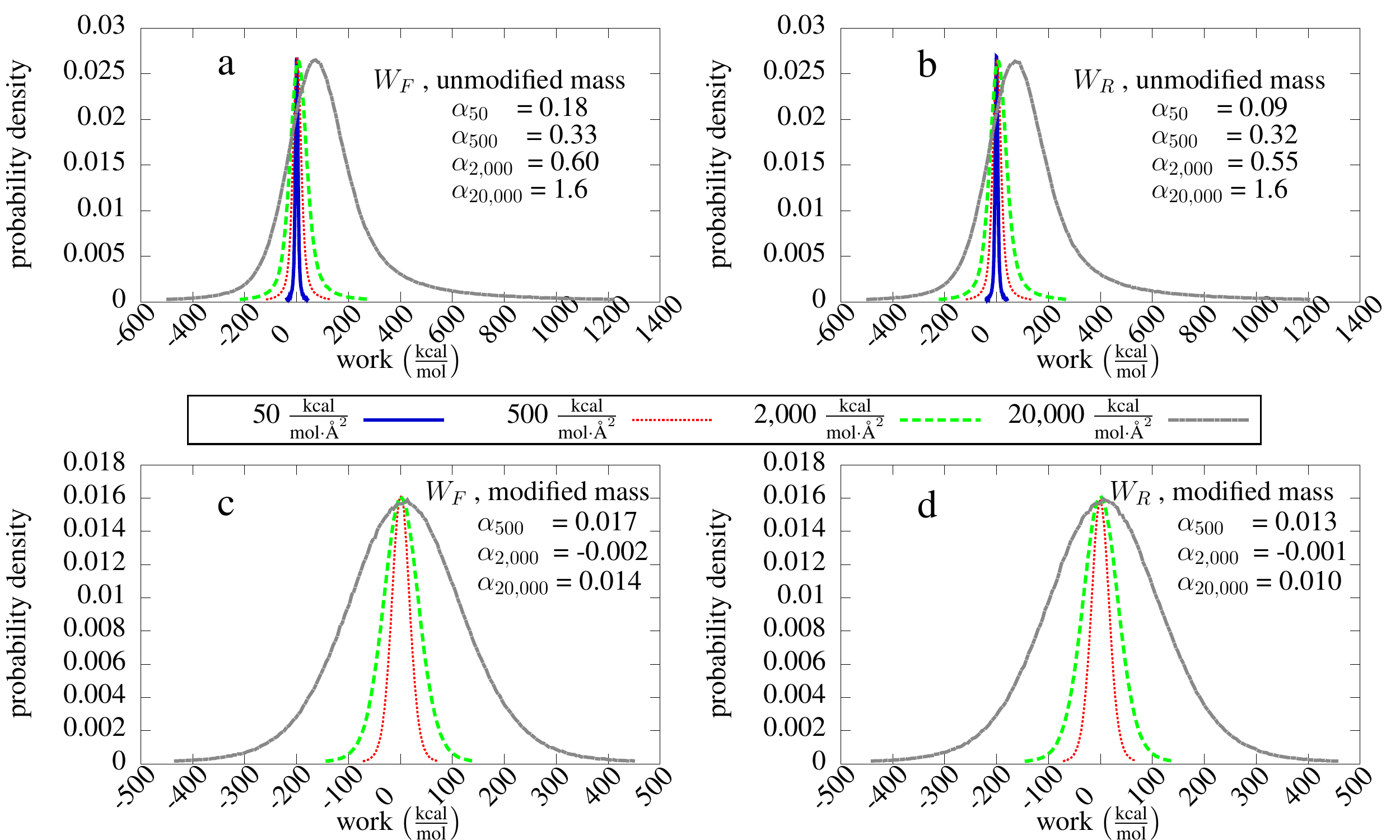}
\caption{K$^+$-Na$^+$ work distributions for the reaction path bin with $z\in[4.0\ang, 5.0\ang]$ from seven series of simulations with different steering spring constants and masses.
The average steering speed (8 \ang/ns) and other simulation conditions are identical among these simulations.
Further simulation details are given in the text.
Work distributions from four series of simulations with spring constants varying by three orders of magnitude are shown in (a) and (b), with forward work (\sod moving away from \pot) in (a) and reverse work (\sod moving toward \pot) in (b).
Similar work distributions are shown in (c) and (d) for three different spring constants, from a system where the masses of the ions are artificially increased to those of the membrane patch and the peptide used in simulations of Fig. \ref{fig_peptide-membrane_distributions}.
It is seen that the trend of increased skewness with increasing the spring constant, seen in (a) and (b), has completely vanished in (c) and (d).
A zooming factor (see earlier text) of 0.01 is used for producing all of the distributions in this figure.
This lack of skewness in (c) and (d) can only be attributed to the change in ions' masses, and exhibits how (under similar conditions) more massive objects are less susceptible to inertial effects, compared to lighter objects of the same geometry and size.}
\label{fig_K-Na_distributions}
\end{figure*}

Work distributions obtained from seven sets of \sod-\pot simulations are shown in Fig.\ref{fig_K-Na_distributions}, for the reaction coordinate bin that extends from x=4.0 \angs to x=5.0 \ang.
These distributions are highly similar to those of other bins, not shown here.
In Figs.\ref{fig_K-Na_distributions}.a and b, work distributions for four systems are shown, where spring constants 50, 500, 2000 and 20000 \kunit have been used to steer the ions.
In Fig.\ref{fig_K-Na_distributions}.a, the work distributions for the forward work (\sod ion moving away from \pot ion) are shown and the corresponding reverse work distributions for the same bin are shown in b.
Aside from slight differences in skewness of the distributions, and an even more minute difference in the location of the forward and reverse work peaks, the distributions are very much the same.
In Figs. \ref{fig_K-Na_distributions}.c and d, work distributions are shown for systems similar to those of a and b, except that the masses of the ions are modified.
The mass of the \pot ion is changed from its natural value of 39.098 u to 92865.78 u, which is the mass of the entire membrane patch used in the simulations for Fig. \ref{fig_peptide-membrane_distributions}.
Similarly, the mass of \sod ion is changed from its natural value of 22.99 u to 1492.83 u, which is the mass of one HHC-36 molecule used in the simulations for Fig. \ref{fig_peptide-membrane_distributions}.

Simulations performed on the ions with these modified masses enable a direct study of the effect of molecules' size and mass on the width and skewness of the work distributions.
Comparison of the work distributions in Figs. \ref{fig_K-Na_distributions}.c and d with those in Figs. \ref{fig_K-Na_distributions}.a and b shows that the width of the work distributions has not been affected by modifying the mass (where every other parameter in the simulation, in particular the spring constant used for steering, was kept the same).
The skewness of the work distributions, however, has nearly vanished by increasing the mass of the steered objects.
For the simulations performed with the stiffest spring ($k$=20,000 \kunit), the skewness has decreased by two orders of magnitude by increasing the mass.

It is instructive also to compare the skewness of the work distributions in Figs. \ref{fig_K-Na_distributions}.c and d with those in Fig. \ref{fig_peptide-membrane_distributions}.c.
The steered objects in the simulations that produced those work distributions, have identical masses, but very different geometries.
The skewness is thus seen also to be a result of the extended geometry of the peptide-membrane system.
When the masses of the peptide and the membrane are incorporated (albeit artificially) in the \sod and \pot ions respectively, using even stiffer springs than those used for the peptide-membrane simulations produces work distributions with essentially no skewness.
We return to the relation between the size and mass of the molecules and the inertial effects in section \ref{fine}.
Skewness of work distributions shown in Figs. \ref{fig_K-Na_distributions}.a and b, together with absence of any skewness in work distributions shown in Fig. \ref{fig_C60-silica_distributions} strongly suggest that the skewness cannot be caused solely by the asymmetric geometry of the system or the steering speed, but is rather much strongly dependent on the inertial effects caused by the use of very stiff steering springs.
Furthermore the fact that the ions used here have no internal degrees of freedom (and yet show skewed work distributions) suggests that the skewness need not be due to energy from the restraint being deposited in vibrational or rotational modes of the solute and being mistakenly registered as work done against the center of mass.

\subsection*{Skewness of work distributions appears only when both steered objects are restrained}
As mentioned earlier in regard with the work distributions obtained from the $\text{C}_{60}$-silica simulations, skewed work distributions are obtained only when the positions of both steered objects (one of which usually held essentially stationary) are restrained using harmonic potentials.
SMD simulations involving more than two objects are, in principle, possible to perform, but we limit our study to the more common case of two-body PMF calculations here.

When one of the steered objects is constrained (held strictly stationary), no skewness is seen in the work distributions.
This was shown by the example of work distributions from the $\text{C}_{60}$-silica system, where the silica slab was constrained, and even using a spring constant of 10,000 \kunit resulted in non-skewed distributions, shown in Fig. \ref{fig_C60-silica_distributions}.
We performed also an SMD FR simulation of the sodium-potasium ions (unmodified masses), with similar conditions as those explained above, with a spring constant of 20,000 \kunit used to steer the \sod ion, and the \pot ion held strictly fixed (constrained) at the origin.
Work distributions from these simulations showed essentially no skewness, but a width similar to those from the simulations with a spring constant of 20,000 \kunit used to steer both ions (data not shown).
Use of a static constraint (holding one of the objects strictly fixed) thus seems to rid us of the skewness problem, but not the broadening of the distributions, caused by using a very stiff spring to steer the other object.
The broadening of work distributions is a major problem in trying to achieve converged PMFs.
It is  often not desirable to use a static constraint on either of the steered objects.
In the case of the peptide-membrane systems, for example, one wishes to capture the various movements within the membrane upon interacting with the peptide, and their effect on the external work (and subsequently, on the PMF).
With a constrained membrane the inter- and intra-membrane movements and interactions are not included in the averaging integral that defines the PMF, and therefore a less informative PMF is obtained, compared to the case when the membrane is restrained.

When a very stiff spring is used to steer both objects, the distribution of work performed on each object alone is (right-)skewed.
This has been verified, e.g., by establishing the work distributions for each of the two ions in the sodium-potassium FR SMD runs with a $k$ of 20,000 \kunit on each ion.
The distribution of external work on each of the two ions was found to be right-skewed, and by about the same value of skewness $\alpha$ found for the total work distributions (data not shown).
The skewness of work distributions, or the underlying inertial effects causing the skewness, is thus a two-body effect.

\section{Inertial effects can result in biased PMF\lowercase{s}}
\begin{figure}[t]
\centering
\includegraphics[width=1.00\linewidth]{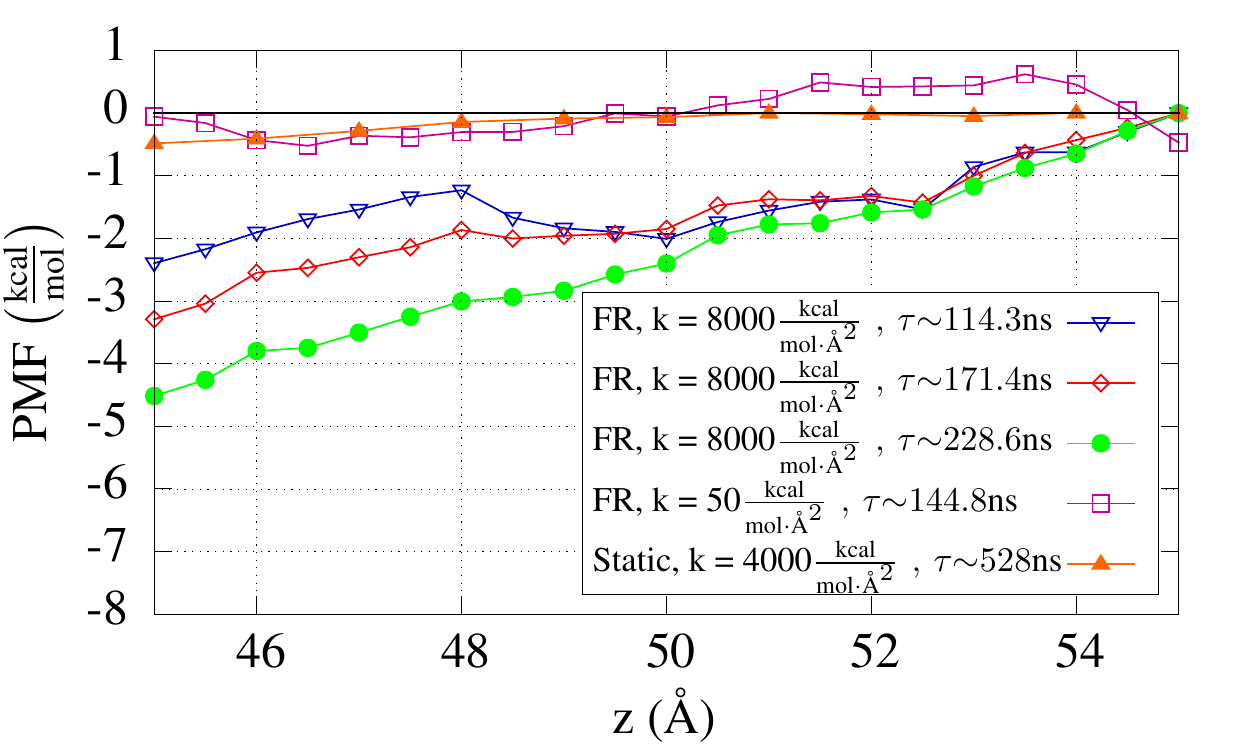}
\caption{PMFs obtained from the same set of SMD simulations of a peptide-membrane system, as described under Fig. \ref{fig_peptide-membrane_distributions} and further in the text, at the far region ($z\in[45\ang,55\ang]$) where there is essentially no interaction and thus a flat PMF is expected.
A static (equilibrium) series of simulations obtains a very flat PMF for the peptide-membrane interaction in this region.
Among the PMFs obtained from SMD simulations, only the one calculated from using the softest spring ($k$ = 50 \kunit) agrees with the PMF obtained using the equilibrium method.
Simulations using a larger spring constant ($k$ = 8000 \kunit) yield sloped (biased) PMFs, and spending more simulation time results in increased bias, as can be seen from the increased slope of the PMFs calculated using longer simulations with the same spring constant ($k$ = 8000 \kunit).
Other than the spring constants, every simulation parameter is the same between the two sets of FR runs shown here.
All the PMFs here are calculated using (\ref{FR}), with average values of $W_F$ and $W_R$ used, as originally prescribed.}
\label{fig_PMFs_bin-passing_far_range}
\end{figure}

\begin{figure}[!]
\centering
\includegraphics[width=1.00\linewidth]{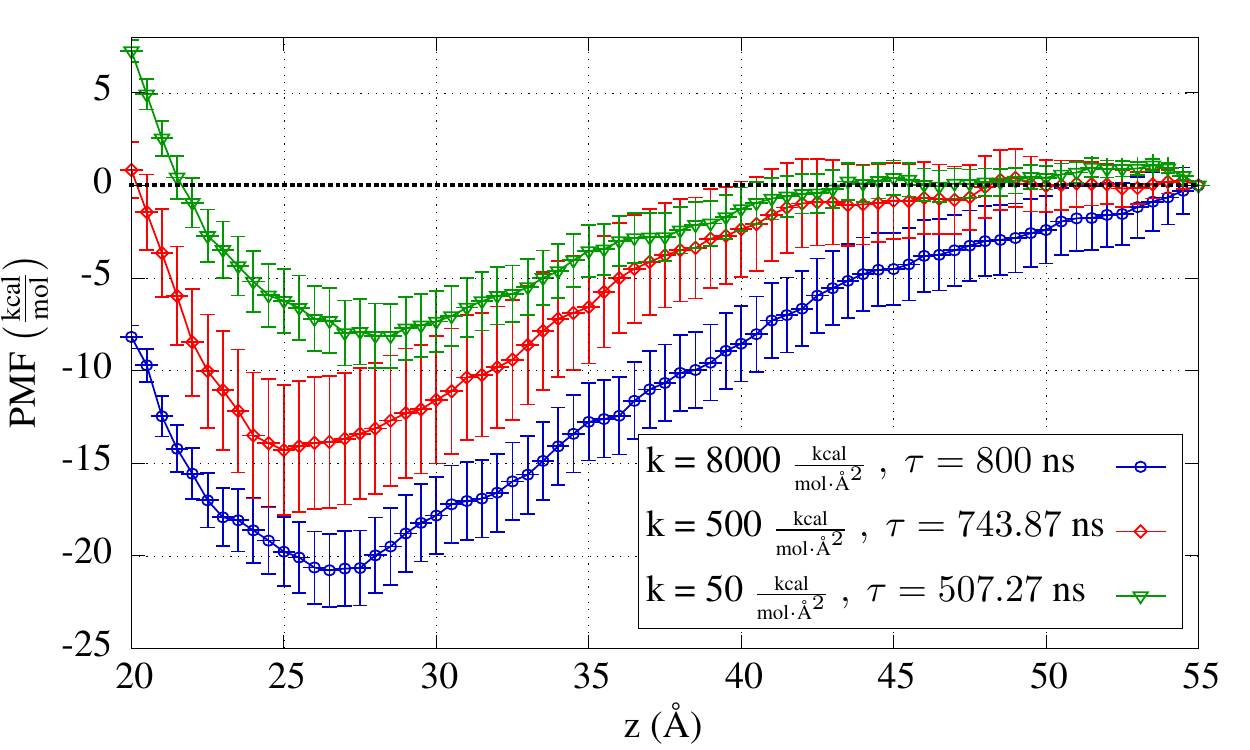}
\caption{PMFs obtained from the same set of SMD simulations of a peptide-membrane system, as described under Fig. \ref{fig_peptide-membrane_distributions} and further in the text.
Note that were it not for the skewness discussed in the text, all three of these PMFs should be the same.
At the far region ($z\gtrsim 40 \ang$) where there is no interaction and thus no slope is expected in the PMF, simulations using larger spring constants yield sloped (biased) PMFs.
All the PMFs here are calculated using (\ref{FR}), with average values of $W_F$ and $W_R$ used, as originally prescribed.}
\label{fig_PMFs_bin-passing}
\end{figure}

In Fig. \ref{fig_PMFs_bin-passing} we show the peptide-membrane PMFs obtained from the whole set of each of the datasets discussed above, but to better assess the effect of the phenomenon shown in Figs. \ref{fig_peptide-membrane_distributions} and \ref{fig_K-Na_distributions} on the PMFs obtained using (\ref{FR}), PMFs are calculated using three different portions of the dataset with the stiffest spring ($k$ =  8000 \kunit) and shown together with the PMF obtained using the softest spring ($k$ = 50 \kunit) in Fig. \ref{fig_PMFs_bin-passing_far_range}.
In the same figure, we also present the PMF obtained from a much longer equilibrium simulation, where static force sampling is used to calculate the PMF.
During these static equilibrium simulations, the peptide is restrained using a spring of stiffness $k$ = 4000 \kunit at 11 locations ($z$ = 45 \ang, 46 \ang , ... , 55 \ang), while the center of mass of the membrane is restrained at $z$ = 0, and the force required to hold the peptide at each location is sampled, and subsequently averaged and integrated to obtain the PMF.
We use this reliable, but computationally much more expensive PMF, as the standard, against which we compare the PMFs obtained using the FR method with different spring constants.
The overshooting phenomenon is not problematic with the static equilibrium method, since only the average of the equilibrated position is used to determine the thermodynamically averaged force, and no accumulated work is required in the calculation.
We have chosen the far-region of the peptide-membrane interaction in Fig.\ref{fig_PMFs_bin-passing_far_range} to demonstrate the bias more clearly, in a region where the correct PMF is flat, as verified by the static equilibrium PMF curve.
Using the data from successively longer portions of the simulation with $k$ = 8000 \kunit is seen to result in increased slope of the PMF curve obtained using (\ref{FR}), while the PMF obtained using the work values from the simulation with $k$ = 50 \kunit exhibits a flatter curve that agrees with the correct PMF obtained using the static equilibrium PMF.
The fact that using longer runs with $k$ = 8000 \kunit worsen the slope of the resultant PMF (and thus the deviation from the equilibrium result), strongly suggests that a systematic error is being increasingly introduced into the free energy profile obtained using the stiffest spring.
This bias, or systematic error in the PMF, can be attributed to the skewness of the work distributions observed in Fig. \ref{fig_peptide-membrane_distributions}, with deviations from the correct value of the resultant average forward and reverse works generally not being equal for each bin.


The PMFs obtained from the simulation with $k = $ 8000 \kunit, shown in Fig. \ref{fig_PMFs_bin-passing_far_range}, exhibit the same type of bias that has been previously reported, e.g. in \cite{Jarzynski_bias1} and \cite{Jarzynski_bias2}.
In the far-from-membrane region, ($z\gtrsim 45 \ang$), where the electrostatic attraction between the cationic peptide  and the anionic membrane surface is effectively screened by water and counterions, one expects to obtain a flat PMF, and this is verified by the static equilibrium PMF, but is clearly not the case for the PMF obtained by using a spring constant of 8000 \kunit.
The repetitive overshootings caused by use of a high spring constant, result in considerable kinetic energy contributions to the external work in both forward and reverse directions (but, in general, not by the same amount).
Average work values obtained from such skewed work distributions result in biased estimations of the PMF.
The bias obtained in the PMFs shown in Figs. \ref{fig_PMFs_bin-passing_far_range} and \ref{fig_PMFs_bin-passing} occurs at a region of the reaction path where the correct underlying PMF is flat.
The observed bias can therefore not be attributed to asymmetries between the forward and reverse works due to the underlying PMF.
It can be seen in Fig. \ref{fig_PMFs_bin-passing} that using smaller spring constants significantly improves the shape and quality of the estimated PMF, in terms of reduced bias (slope in the far region) as well as the reduced size of the estimated error-bars for the $k$ = 50 \kunit PMF.
The PMF obtained using a spring constant of 500 \kunit falls between those obtained spring constants of 8000 \kunit and 50 \kunit, as anticipated from the behavior of the corresponding work distributions in Fig. \ref{fig_peptide-membrane_distributions}.
This further supports the connection between the bias in the PMF calculated from a set of non-equilibrium trajectories and the skewness of its work distributions.

\section{Fine-tuning the stiff spring approximation}
\label{fine}

The proper value of the steering spring constant $k$ to be used in SMD simulations, can be found using two criteria.
The upper limit for $k$ is found by requiring that the system remains in the overdamped limit of the Langevin equation.
This is indeed a requirement of SSA and thus for the FR method, but the necessary conditions for its validity can be overlooked in design of simulations, as using stiffer steering potentials might be desirable for more precise steering.
If the steered object is considered as a damped harmonic oscillator subject to a steering force $-k~ x$ and a velocity dependent friction force $-\gamma \dot{x}$, Newton's second law reads

\begin{equation}
m \ddot{x} + \gamma \dot{x} + kx = 0
\label{Newton}
\end{equation}
the overdamped limit of this equation occurs when $\tfrac{\gamma}{2m} > \sqrt{\frac{k}{m}}$.
For typical biomolecular simulations, where water often constitutes the surrounding environment, and for reasonably small steering velocities (say, 10 \ang/ns, appropriate for biomolecular SMD simulations), the Reynolds number is small and the value of $\gamma$ for the steered molecules can be estimated from Stokes' law
\begin{equation}
\gamma = 6 \pi \eta r,
\label{Stokes}
\end{equation}
wherein $\eta$ is the shear viscosity of the fluid, and r is the radius of the steered object (approximated to have a spherical shape).
Substituting (\ref{Stokes}) in the condition for the overdamped limit and solving for $k$, we obtain
\begin{equation}
k < \frac{9 \pi^2 \eta^2 r^2}{m}
\label{k up}
\end{equation}
It might seem that (\ref{k up}) gives an upper limit for $k$ that decreases with increase in size of the molecule, as one might expect $m$ to be roughly proportional to $r^3$, which might lead to a $\sim 1/r$ dependence for allowed upper limit of $k$.
However, molecules involved in SMD simulations are often polymers (e.g., proteins), whose radius of gyration depends, in general, on their number of monomeric units ($N$) roughly as $R \sim N^\nu$, where the value of the scaling exponent $\nu$ depends on the solvent surrounding the polymer.
In a good solvent, $\nu$ can be as high as 0.59, whereas in a bad solvent, $\nu$ can reach the value 1/2 \cite{rubinstein2003polymer}.
In SMD simulations, the molecules steered to move are usually soluble in the solvent, so it is safe to assume a $\nu$ of larger than 1/2, which (considering that $m \sim N$), gives a power law dependence on $r$ in the denominator of the right-hand-side of (\ref{k up}) with the exponent less than two.
The dependence of upper limit of $k$ on $r$ would thus be a power law with a positive exponent, meaning that $k$ increases with $r$.

The lower limit for $k$ is simply found by requiring that the restraining potential energy $\tfrac{1}{2} k (\delta x)^2$ remains larger than the average thermal kinetic energy per degree of freedom in the system $\tfrac{1}{2} k_B T$ \cite{FR}. Solving for $k$, this gives
\begin{equation}
k > \tfrac{k_B T}{(\delta x)^2}
\label{k low}
\end{equation}
Here $\delta x$ represents the maximum tolerable deviation of the steered object from the prescribed trajectory.
For most biomolecular simulations, a sub-Angstrom value is often a proper value for $\delta x$.

For the HHC-36 peptide with a mass of 1.48 ku, and for simulations at T = 310 K, when a precision of 0.1 \angs is sought, (\ref{k up}) gives an upper limit of 24.7 \kunit, when (\ref{k low}) gives a lower limit of 61.6 \kunit.
As these limits give no common range, a value of 50 \kunit has been chosen and used to meet both overdamping and precision requirements as closely as possible.
It is worth noting that (\ref{k up}) has a size-dependence, while (\ref{k low}) depends only on temperature and the sought precision.
The permitted upper limit for $k$ given by (\ref{k up}) increases slowly (perhaps more slowly than linearly) with the size of the steered objects, and this poses a problem to SMD simulations of some large molecules, where the upper limit permitted by (\ref{k up}) may not have a common range with the limit determined by (\ref{k low}).
In the next section we see how the peak-finding method can offer a remedy.

\section{The peak-finding method}
\begin{figure}[!]
\centering
\includegraphics[width=1\linewidth]{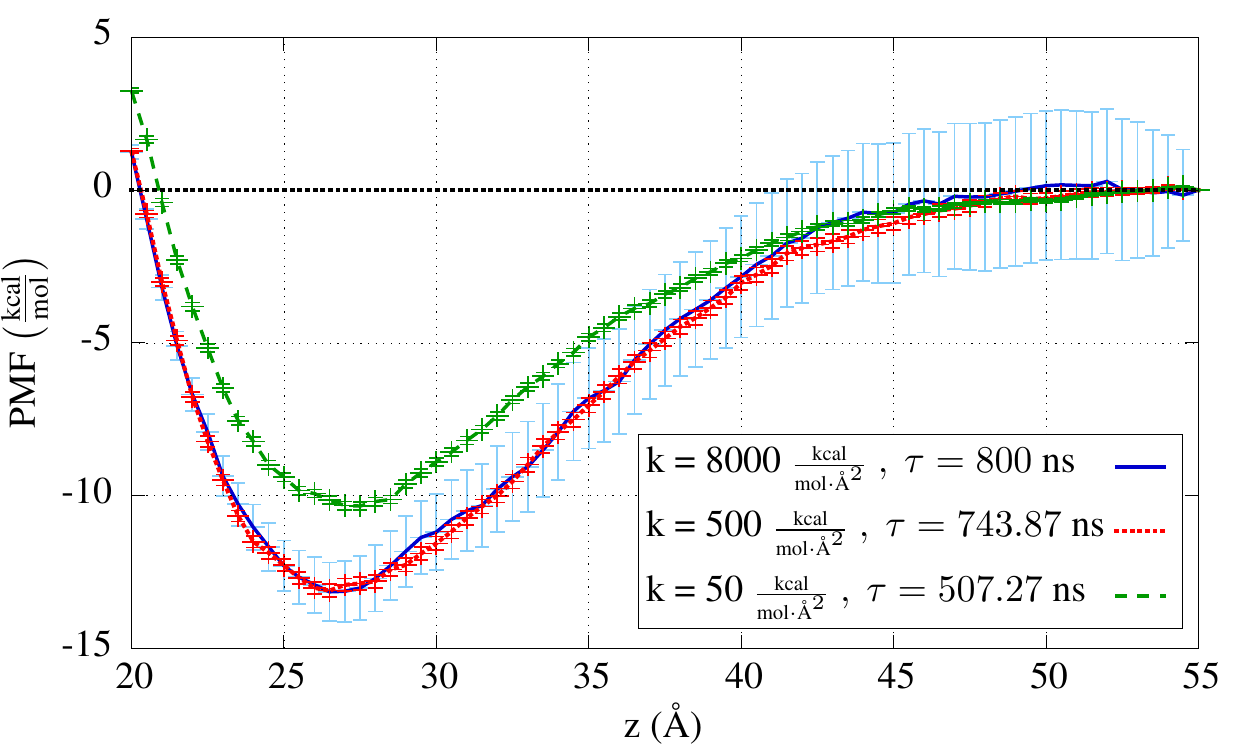}
\caption{PMFs for the interaction of the HHC-36 peptide with a POPE/POPG membrane patch, obtained using the \textit{peak-finding} method.
The same simulation trajectory and data is used as in Fig. \ref{fig_PMFs_bin-passing}.
Use of peak values in (\ref{FR}) has resulted in less biased PMFs, even when a $k$ of 8000 \kunit has been employed.
Here $\tau$ is the overall simulation time in each case.
For smaller $k$ values, the peak-finding method obtains remarkably smaller error-bars (shown in the figure, but often smaller in size than the data-point symbols for the $k$ = 50 \kunit curve).}
\label{fig_PMFs_peak-finding}
\end{figure}

The bias in the PMFs obtained from SMD simulations, introduced by the inertial effects, can be removed or at least greatly reduced by using softer springs, and observing (e.g., in Fig. \ref{fig_peptide-membrane_distributions}) that even when the work distributions exhibit a deviation from the Gaussian shape, the peak-location of the distributions does not vary as strongly with $k$ as does the mean value.
When work distributions are skewed and $\langle W \rangle$ is biased, larger work values become more probable (thus the right-skewness), but the peak-location of the work distribution is not as significantly shifted as the mean value, as seen in Figs. \ref{fig_peptide-membrane_distributions}.b, c and d.
The peak location of the work distribution can thus be used as a better estimate of the unbiased $\langle W \rangle$.
This requires accurate estimation of $W_{peak}$ in each direction and for each reaction path bin, to be used in (\ref{FR}) in place of $\langle W \rangle$.
Fig. \ref{fig_PMFs_peak-finding} demonstrates the improvement obtained by following such a procedure, where the same data-sets demonstrated in Fig. \ref{fig_peptide-membrane_distributions} and used to obtain the PMFs in Fig. \ref{fig_PMFs_bin-passing} are used, but $W_{peak}$ is used in place of $\langle W \rangle$ in (\ref{FR}).
Even for the most skewed work distributions, i.e., those with $k$ = 8000 \kunit, this method gives a much less biased estimate of the PMF.
For simulations done with $k$ of 500 and 50 \kunit, this \textit{peak-finding} method provides smoother PMFs with remarkably smaller error bars, shown in Fig. \ref{fig_PMFs_peak-finding}.

\begin{figure*}[!]
\centering
\includegraphics[width=1.0\linewidth]{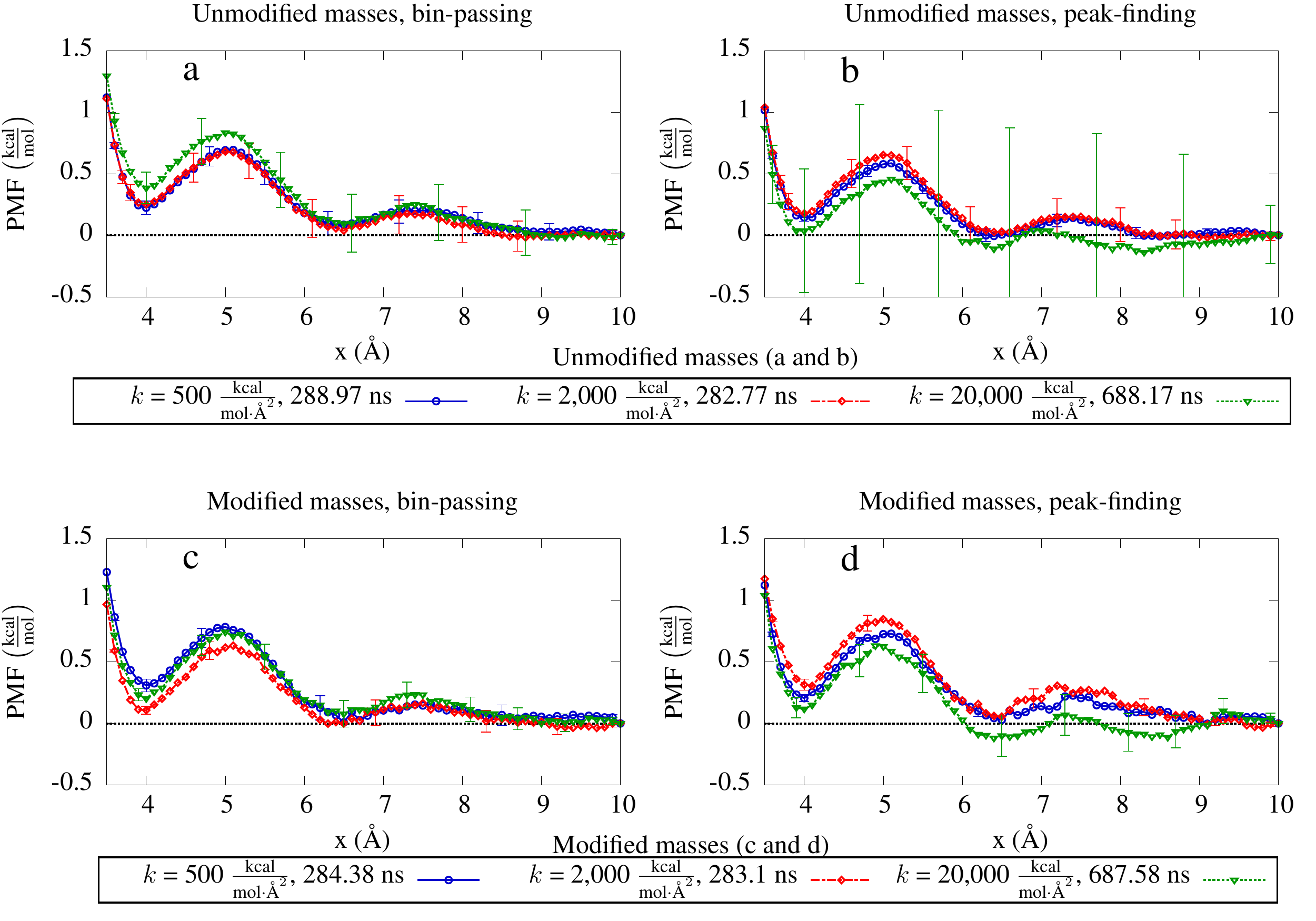}
\caption{K$^+$-Na$^+$ PMFs calculated using bin-passing (a and c) and peak-finding (b and d) methods, for systems where the ion masses are left at their natural values (a and b) and modified (c and d).
To better view the PMF curves, only some of the error bars are shown, whose size is nevertheless representative of the average size of error bars in their respective neighboring regions.
Simulation times are shown in the legends.
Results obtained from simulations with the stiffest spring ($k$ = 20,000 \kunit) produce the largest error-bars, despite those simulations being run for over twice as long as the others.
The size of error-bars in this figure can be seen to correlate with the width of corresponding work distributions in Fig. \ref{fig_K-Na_distributions}.
The skewness of work distributions has resulted in biased PMFs, but to a lesser extent than in the case of the peptide-membrane system examined earlier (compare the green curve with the red and blue curves in a).
The best PMFs (assessed by pairwise agreement and size of error-bars) here are the ones obtained from the system with the unmodified mass and with the softer springs (red and blue curves in a and b).}
\label{fig_K-Na_PMFs}
\end{figure*}

As another example of the application of the peak-finding method, we consider the \pot-\sod interaction PMF in water, discussed earlier.
For the six series of simulations performed on this system (all sharing the same steering speed and other general properties given earlier) with two different sets of mass values (natural and modified) and three different steering spring constants of 500, 2,000 and 20,000 \kunit, the PMFs are given in Fig. \ref{fig_K-Na_PMFs}.
The lower limit for $k$ in this system (with T = 310 K and assuming a $\delta x$ of 0.1 \ang), as given by (\ref{k low}) is 61.6 \kunit.
Use of equation (\ref{k up}) for the \sod ion in this system gives an upper limit of 21.6 \kunit for $k$.
However, for SMD simulations involving such small ions as \sod and K$^+$, much higher values for $k$ can safely be used (here we used a $k$ of 500 \kunit as the lowest spring constant) without introducing significant systematic error to the calculated PMF.

For each series of simulations (with a given mass and spring constant), PMFs calculated using both the bin-passing (averaging) \cite{bin-passing} and peak-finding methods are shown.
For this system too, the skewed work distributions, have resulted in a biased (sloped) PMF: in the simulations with unmodified (natural) masses and with the highest spring constant ($k$ = 20,000 \kunit), it was shown in Figs. \ref{fig_K-Na_distributions}.a and b that the resulting work distributions were highly skewed.
That skewness has resulted in deviation of the resultant PMF by a maximum of $\sim$ 0.8 kcal/mol, from the PMF obtained using the softest steering spring ($k$ = 500 \kunit).
While this difference is within the error bars of the two curves in some regions, the curve obtained using $k$ = 20,000 \kunit deviates from that obtained by $k$ = 500 \kunit, with the difference increasing from high distances to low distances, suggesting the presence of bias.
Smaller bias in the PMF obtained using the stiffest spring here (in the system with unmodified mass), compared to the peptide-membrane system, is very likely the result of  less asymmetry between forward and reverse pulls and the absence of the extended geometry of a peptide-membrane system.
The features of the \pot-\sod PMF are also much finer than the peptide-membrane PMF considered above, putting a higher precision demand on the PMF calculation method.
The skewed distributions result also in poorer convergence of the calculated PMF.
The PMFs obtained using both the bin-passing (averaging) and peak-finding methods from simulations with the stiffest spring on the system with the modified (increased) masses of ions, exhibit better convergence than those with the same spring constant on the system with the unmodified masses.
This is despite the approximately similar width of work distributions between the two systems.
In the system with unmodified masses the work distributions are highly skewed, which has resulted in larger uncertainty in calculating both the average and the peak-value of work distributions, and subsequently, the PMFs.

Implementing the \textit{peak-finding} method is potentially challenging, as one is required to first establish the work distributions for each bin along the reaction path in both forward and reverse directions and then to reliably and accurately determine the peak-location of each distribution.
Doing this, e.g., for each of the PMFs shown in Fig. \ref{fig_PMFs_peak-finding}, requires establishing 140 distributions (130 distributions for each PMF shown in Figs. \ref{fig_K-Na_PMFs}.b and d), finding their peak locations and also performing the proper error analysis.
To establish the distributions in a way that is useful for subsequent analysis, we perform the \textit{zooming} procedures described in previous sections and demonstrated in Fig. \ref{fig_peptide-membrane_distributions} with a high value of the zooming factor $f$.
A value of 0.75 was used for $f$ in obtaining the distributions used to calculate the PMFs in Figs. \ref{fig_PMFs_peak-finding} and \ref{fig_K-Na_PMFs}.b and d.
The zooming procedure is done using the \textit{bidirectional work distribution builder} \cite{software_work-distribution-builder} software that we have created for this purpose and made publicly available.
Using this code, each of the distributions shown in Fig. \ref{fig_peptide-membrane_distributions} can be obtained from the simulation raw data (i.e., record of external forces and object locations during each time-step).

One can then in principle use a non-linear fitting algorithm to locate the peak of the distributions such as those shown in Fig. \ref{fig_peptide-membrane_distributions}.c.
Using non-linear fitting methods, however introduces some arbitrariness, as it often requires good initial guessed values for the fitting parameters and thus in general will not obtain exactly reproducible results.
We have instead devised a linear method as follows: we use the \textit{bidirectional work distribution builder} \cite{software_work-distribution-builder} software to zoom into the peak-location of each given work distribution with a high zooming factor ($f$), as shown in parts b, c and d of Fig. \ref{fig_peptide-membrane_distributions}.
We found a zooming factor of 0.75 appropriate for linear peak-finding procedure, and it gives distributions such as those shown in Fig. \ref{fig_peptide-membrane_distributions}.d with one run of the code.

Such distributions, which essentially exhibit the peak of each work distribution, can be conveniently fitted to a quadratic polynomial, using a linear least-square fitting algorithm.
If the coefficients $a$, $b$ and $c$ of a quadratic polynomial $ax^2 + bx + c$ are adjusted to fit a distribution of the form given in Fig. \ref{fig_peptide-membrane_distributions}.d, the peak value of the distribution will simply be given by $-b/(2a)$, with an associated uncertainty $\delta W_{peak}$ equal to $\sqrt{(\delta a/a)^2 + (\delta b / b)^2}$, where $\delta a$ and $\delta b$ are the uncertainties in finding coefficients $a$ and $b$, respectively.
This is conveniently done using another software created by our group, the \textit{work distribution peak finder} code \cite{software_work-distribution-peak-finder} that  utilizes the linear least-square fitting libraries provided by the GNU scientific library (GSL) \cite{gsl}.
The \textit{work distribution peak finder} software can fit the given distributions (produced by the \textit{bidirectional work distribution builder} software) to polynomials of order two or higher, but we have found that using polynomials of orders higher than two will not improve the accuracy of the PMF calculated from the resultant $W_{peak}$ values.
Furthermore, the uncertainty in the calculated  $W_{peak}$ is most easily found when a quadratic polynomial is used for fitting, which vastly simplifies the task of finding the uncertainty of the PMF calculated from those $W_{peak}$ values using (\ref{FR}).
To obtain the PMFs shown in Fig. \ref{fig_PMFs_peak-finding}, a quadratic polynomial fitting algorithm has been employed.
The \textit{bidirectional work distribution builder} \cite{software_work-distribution-builder} and \textit{work distribution peak finder} \cite{software_work-distribution-peak-finder} codes together enable calculating PMFs of very high accuracy and with very small bias from SMD trajectories of large biomolecular systems.
Both softwares are provided under the terms of the GNU general public license (version 3).

\section{Conclusions}
The bias introduced into the PMFs found using NEW theorems for large biomolecular systems was shown here to have its root in the skewness of the work distributions obtained from SMD simulations.
The skewness, in turn, is caused by inertial effects due to use of too stiff guiding potentials.
We found the main cause of the inertial effects to be the improper (too large) value of the spring constant $k$ (for a given mass of the steered objects) and provided the relations for choosing $k$ such that the inertial effects can be avoided or minimized.
We examined three other possible sources of the skewness of work distributions, namely the steering speed, mass and the physical asymmetry in the systems.
In absence of too high $k$ in SMD simulations, no skewness was seen to be caused by physical asymmetry, high mass, or high steering speeds, within the range of systems and speeds we studied.
However, when inertial effects are present (and the work distributions are thus skewed), physical asymmetry in the simulated system seems to translate into asymmetry between the skewnesses of forward and reverse work distributions, resulting in biased PMFs.
Physical asymmetry thus seems to be a contributer (but not a sufficient cause by itself) to introduction of bias into the PMFs calculated from SMD simulations.
This can be seen by comparing the results obtained for the peptide-membrane and \pot-\sod systems: while skewness of work distributions is higher in the simulations with highest $k$ in the \pot-\sod system, the bias in the calculated PMF is much higher in the peptide-membrane system (also when $k$ is at its highest).
When $k$ is not too high, physical asymmetry causes neither skewed work distributions nor biased PMFs in SMD simulations.

In absence of inertial effects, one expects very nearly Gaussian work distributions obtained from SMD simulations, as predicted by SSA.
Departure from that shape, in the form of right-skewed work distributions, is a consequence of inertial effects.
Using such work samplings in any of the currently available NEW theorems, inevitably results in systematic errors introduced into the calculated PMFs.
Relations (\ref{k up}) and (\ref{k low}) can give safe limits for the value of $k$ to be used in a given system, thereby avoiding the pathology just described.
In any case, using peak-values (rather than averages) of the external work distributions in (\ref{FR}) results both in reduction of the bias, and considerable increase in the accuracy of the calculated PMFs.
We provided the physical justification as well as the procedure for calculating the PMFs from the peak of work distributions and the software for building the work distributions \cite{software_work-distribution-builder} and obtaining the PMFs from their peak values \cite{software_work-distribution-peak-finder}.

The pathology discussed here, is harder to spot and quantify in smaller systems, for which short simulations ($\lesssim$ 100 ns) can obtain converged PMFs using NEW methods.
This was the case for the \pot-\sod system we studied here.
As is often the case, this bias surfaces when longer samplings are needed to achieve convergence, resulting in accumulation of the systematic error.
If the resultant biased PMFs are then used for calculating other quantities such as radial distribution function or diffusion coefficients, this bias will then propagate.
Care should thus be taken in designing SMD simulations, and also in interpreting the PMFs obtained from them.
\begin{acknowledgements}
The authors are grateful to the Natural Sciences and Engineering Research Council of Canada for financial support.
Computational resources for the simulations presented in this work were provided by Calcul Qu{\'e}bec HPC consortium of Compute Canada.
\end{acknowledgements}

\providecommand{\noopsort}[1]{}\providecommand{\singleletter}[1]{#1}%

\end{document}